# Explainability via Interactivity? Supporting Nonexperts' Sensemaking of Pretrained CNN by Interacting with Their Daily Surroundings

Chao Wang

Honda Research Institute Europe, Offenbach/Main, Germany, chao.wang@honda-ri.de

Pengcheng An

David R. Cheriton School of Computer Science, University of Waterloo, Waterloo, Ontario, Canada, anpengcheng88@gmail.com

Current research on Explainable AI (XAI) heavily targets on expert users (data scientists or AI developers). However, increasing importance has been argued for making AI more understandable to nonexperts, who are expected to leverage AI techniques, but have limited knowledge about AI. We present a mobile application to support nonexperts to interactively make sense of Convolutional Neural Networks (CNN); it allows users to play with a pretrained CNN by taking pictures of their surrounding objects. We use an up-to-date XAI technique (Class Activation Map) to intuitively visualize the model's decision (the most important image regions that lead to a certain result). Deployed in a university course, this playful learning tool was found to support design students to gain vivid understandings about the capabilities and limitations of pretrained CNNs in real-world environments. Concrete examples of students' playful explorations are reported to characterize their sensemaking processes reflecting different depths of thought.

CCS CONCEPTS • Human-centered computing~Ubiquitous and mobile computing~Ubiquitous and mobile computing design and evaluation methods

## 1 INTRODUCTION

With its advantages of solving computer vision problems such as object detection or image classification, Convolutional Neural Networks (CNN) are wildly applied in various domains e.g., autonomous driving, healthcare and robotics. Especially after the AI community open-sourced frameworks such as TensorFlow[1], Caffe[2] or PyTorch[3], non-expert users, such as user experience designers, healthcare practitioners or even AI amateurs, could affordably utilize, or adapt a pretrained CNN model (e.g., VGG-19 [4]) for their own pragmatic purposes, which is expected to democratize CNN models for real-world usage. However, CNN is a complex type of deep learning models which normally consist of multiple convolutional, pooling and fully connected layers. For non-expert users who have limited knowledge about AI, it is rather difficult for them to make sense of how a pretrained CNN works and what is its capabilities and limitations in real-world contexts. This creates obstacles for them to utilize pretrained models as "building blocks" [5] to create their own domain-relevant applications. Therefore, non-experts were often baffled when model performance did not fit their expectations, and hence might abandon their tasks [6]. Nowadays, machine learning experts can use various Explainable Artificial Intelligent (XAI) approaches, such as Class Activation Maps [7] or Deconvnet [8] to interpret the internal state of the CNN and reason the characteristics of the trained model. But such techniques are mainly developed for professional machine learning engineers, whose knowledge level and pragmatic goals differ from non-expert users. Aiming to explore how to help non-experts to intuitively understand pretrained CNN models, we design a playful tool that allows users to apply a CNN model to interact with their daily surroundings. This application has been deployed in a university course to help 30 design students (who have zero or little experience with machine learning) to interactively make sense of CNN and understand its capabilities and limitations. We see the application as a technology probe to surface how non-experts' sensemaking process could be scaffolded, and gather implications for future design of similar tools. To address preliminary findings, we detail a few students' experiences of playing with the tool in their daily environment, as well as typical examples reflecting how their sensemaking processes were supported. We uncover three types of sensemaking activities that could be supported by the application, which differed in the depth of thought: *collection* of "impressive" or "funny" results, *inferences* of how environmental factors influenced the results, and *experimentation* on-the-spot to test assumptions. We thereby contribute a playful tool, as well as empirical implications of helping nonexperts to gain practical understandings of pretrained CNN models, via interacting with their daily surroundings.

## 2 RELATED WORK

### 2.1 XAI for non-expert users

There is not a unitary definition of Explainable Artificial Intelligence (XAI). As Arrieta et al. [1] put, "Given an audience, an explainable Artificial Intelligence is one that produces details or reasons to make its functioning clear or easy to understand". From this definition, one crucial consideration of XAI would be the audience, as the purposes of the explanation can vary with various audience. However, most of the existing XAI approaches, by default, target on expert users: e.g., AI specialists or data scientists. Increasingly, researchers point out the necessity of supporting non-experts to practically understand AI models[6][5]. Our approach focuses on such non-experts (namely designers), who are expected to benefit from utilizing pretrained AI model in their domain-specific tasks [9][10][11], but only have limited knowledge about AI.

### 2.2 Approaches of enhancing explainability of CNN

Regarding the HOW in XAI, Guidotti et al [12] and Arrieta et al [13] suggested a general distinction between transparent models and post-hoc explainability. Post-hoc explainability targets models that are not readily interpretable by design, also called black-box models. One of the most focused topics of black-box XAI is to help people understand Convolutional Neural Network (CNN), a widely used model for computer vision problems (e.g., image classification or object detection). CNN consists of a sequence of convolutional layers and pooling layers to automatically learn features incrementally from low to high levels. The functioning and architecture of CNN is extremely complex and difficult to explain. Fortunately, the road to explainability for CNN has become much easier, after researchers found the Activation Map approach, which leverages human brains' intrinsic skills of processing visual data [13]. This approach visualizes the decision process of CNN by propagating the output to the input space to depict which parts of the input were "important" for the output (see Figure 1). This approach has been investigated by researchers extensively. One of the seminal works is proposed by Zeiler et al [8]., who tried to use Deconvnet [14] to reconstructs the maximum activations. Through a saliency map, human can get an intuitive image about which parts of the image significantly contribute to the activations. However, these methods are not class discriminative. Later, Zhou et al. modified CNN though global average pooling [7] to generate class activation maps (CAM), which indicates the discriminative image regions used by CNN to identify the prediction [15]. Based on their work, Selvaraju et al. [16] further proposed Gradient-weighted Class Activation Mapping (Grad-CAM), which is also able to show important part of the input respect to the prediction, but without any modification in the network architecture. In general, highlighting the most discriminative area upon the input image is an important approach for explain and evaluate a pretrained CNN model. Inspired by these works, our solution utilizes a similar approach to visualize the activation of CNN to more intuitively show non-experts how the classification results are generated.

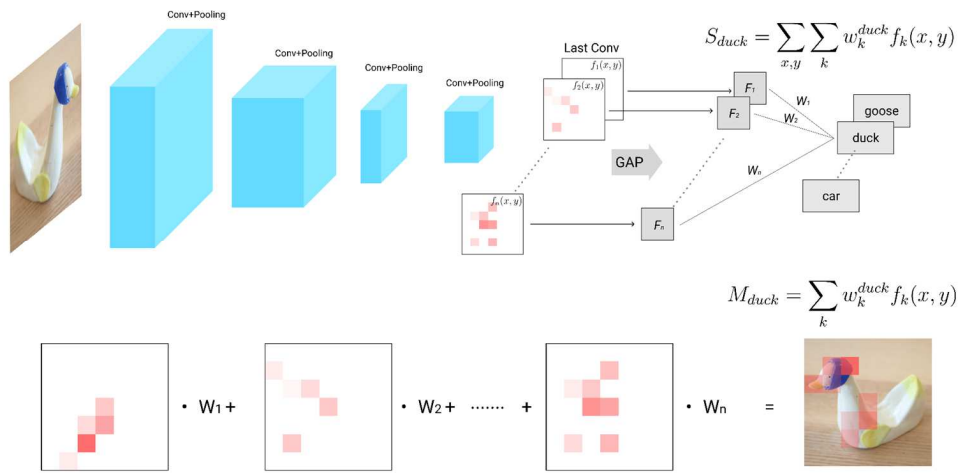

Figure 1 Visualizing discriminative regions with Class Activation Mapping [15]. $f_k(x, y)$ represents the activation of unit k in the last convolutional layer at spatial location (x, y). $F_k$ is the result of performing global average pooling for the unit k. Then, for a certain prediction result, $w_1$, $w_2$ …$w_n$ are the weights for $F_k$ to calculate the softmax input S (e.g., $S_{duck}$). As $F_k$ comes from the global averaged pooling, the corresponding weight also indicates the importance of unit k. Thus, the aggregated heatmap ($M_{duck}$) reflects the activation of the last convolutional layer, hence indicating the most important regions (red squares) that have made the CNN to output a certain prediction result (e.g., duck).

### 2.3 Platforms for XAI of CNN.

Although the topic of XAI for CNN is gaining increasing attention, most the of the tools are built for machine learning experts. Few platforms help non-expert users to develop practical and intuitive understandings about CNN. Yet one example is from the PAIR team of Google, called Embedding Projector [17], which can visualize the features of CNN model in a 3D space after reducing dimensions by T-SNE. Input images of various layers are laid out in a 3D array in its GUI, to illustrate how the input is processes layer by layer. Although granting a global view of CNN, it does not mean to explain how a prediction is generated. Mishra et al. [5] proposed an interactive transfer learning tool for non-experts. User can drag and drop a grey square on any part of an image and observe the influence of the occlusion on the prediction result. Another similar study is ClickMe.ai, a large-scale online experiment conducted by Linsley et al [18]. Authors made an interface for human participants to select important image parts with their mouse. Meanwhile, these parts are shown to CNN incrementally until the network recognize the object. Later, Demidov [19] created a website, which allows user to input an online image to see a corresponding activation heat map through CAM. Although these works inspired us to explore XAI through interactivity, they were all developed for desktop environments and used online images as input source. By contrast, we present a mobile tool that helps users understand CNN's capabilities and limitations by freely interacting with their real-world surroundings.

## 3 SYSTEM DESIGN

Our mobile application enables non-expert users to take photos of surrounding objects and visualizes the CNN's decision about a classification result using a Class Activation Map approach. Namely, it highlights the discriminative image regions (i.e., the red blocks as seen in Figure 1), to intuitively show users which regions of the input image have made the model to decide that the image is about a certain object class. To guarantee the ease of access, we built the tool using a web-based machine learning toolkit, TensorFlow.js [20], so that it does not require any installation on users' mobile device. A pretrained CNN model, MobileNet [21], is integrated in the tool, which is trained upon the ImageNet database (with 10 million labelled images) to recognize over 1000 object categories. This enables users to open-endedly apply the CNN with different kinds of daily objects, and have meaningful recognition results for them to make sense of the CNN's functioning in real-world settings.

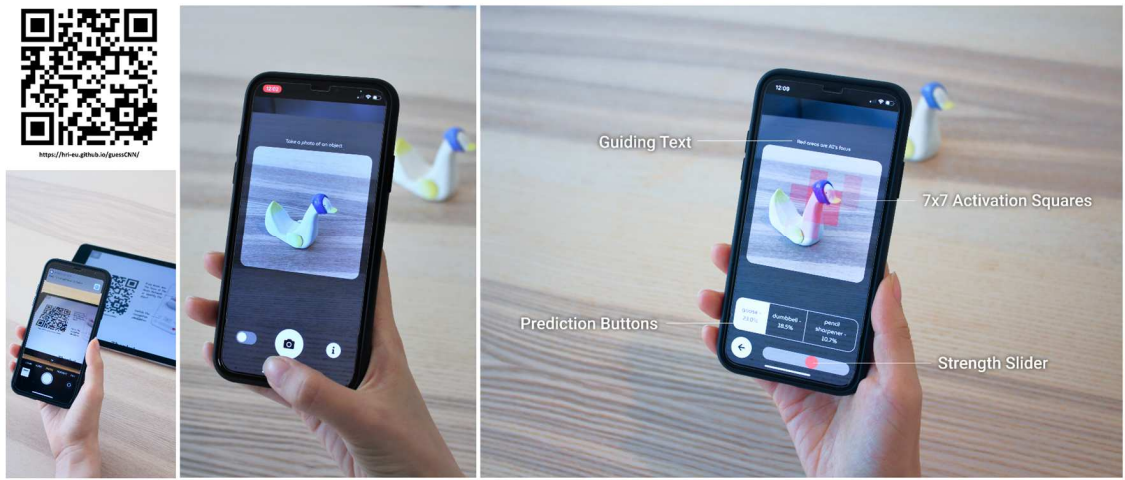

Figure 2. Left: log in via the QR code (currently available). Middle: user takes photo of an object. Right: afterwards, user can see the predictions (three results with highest confidence) and their corresponding Class Activation Maps.

### 3.1 System interaction

Users can visit the URL address (https://hri-eu.github.io/guessCNN/) or simply scan the QR code (see Figure 2, currently available) to access our tool. At the beginning, users need to give the permission for the application to use the back camera. At the same time, the pretrained CNN model is downloaded with the progress shown. After the model is downloaded, the buttons are turned to available status. When clicking "i" button on the bottom right, an overlay page will pop-up to show the introduction of the application and a simple tutorial through a picture; User can aim to an object and take a picture by clicking bottom middle, then the app will start applying the CNN to recognize the taken picture (see Figure 2 Right). Via the toggle button on the bottom left, user can go back to the camera mode to take new pictures.

### 3.2 Generating Class Activation Map

The interface will show top 3 prediction results (with highest confidence) from the overall output by the pretrained model (Figure 2 Right). The model was modified through Class Activation Map (CAM) approach to generate heatmap (Figure 1). Namely, the last fully connected layer of the CNN is replaced by the average pooling layer. 7x7 red squares indicate the discriminative image regions which contribute most to the result of the classification. Users could toggle among the 3 results to see corresponding CAM. A slider on the bottom can modify the thresholds of the activation map display, to adjust the relative intensity and visibility of the squares.

## 4 STUDY SETUP

The interface was implemented as part of a university course module named "prototyping with AI", involving 30 students (9 master students, 21 bachelors) in an industrial design department. Before this module, these students did not take courses encompassing the topic of artificial intelligence, and they had no prior experience with developing or implementing CNN models. This interface was meant as a playful tool for the design students to get familiarized with the pre-trained CNN model, so that later they were prepared to apply a transfer learning technique to modify a CNN model to perform their own customized tasks (using the Teachable Machine [22] platform developed by Google). Our mobile application were introduced to the students in the beginning of the module via an online video lecture (because of the covid-19 pandemic). This video included examples of how to use this interface. After the introduction, an assignment was given in the video, which asked the students to use this application to interact with their daily surroudnings and try to practically make sense of the functioning of the CNN model in their own environment. The students were encouraged to record important observations via screenshots, and share interesting examples with each other. To gather empirical implications, three students participated in in-depth interviews to talk about their experience with the application, especially concrete examples of how they used the application to interact with their daily surroundings and make sense of the classification results. These interviews were subjected (along with the related screenshots sent by the students) to an affinity diagram analysis.

## 5 FINDINGS AND IMPLICATIONS

In general, the students had rather positive experience with the application, which they considered to be both fun to play with and useful for understanding how the CNN model works. We now further address their experience and sense-making process using concrete examples identified by themselves, which have been categorized into three main clusters. These examples illustrate how students' sensemaking of the CNN model could be supported in different depths. We refer to the interviewed students as Jill, Ada, and Sam.

**Collection of "impressive" or "funny" classification results in the surrounding**. The students all mentioned that they enjoyed the playful process of exploring the mobile application with various daily objects, and collecting interesting cases in which the classification results seemed extra impressive or funny to them. They felt that interacting with surrounding objects provided them with more vivid understandings of the model's capabilities and limitations in real-world situations. For instance, one of the examples collected by Jill is a remote (TV) control which was correctly classified by the model with 100% confidence (see Figure 3). In the introductory lecture, it was mentioned that the web interface could classify certain internet images with 100% confidence, because the image might be part of the training dataset of the model. Therefore, it was noticeable to her that a randomly taken picture of her remote control could also reach 100% confidence. Apart from such impressive classification results, the students also enjoyed collecting the funny results. For example, Sam shared a case that the model classified an image of a chair on his book cover as a plunger (56.5% confidence). This feels quite funny to him, but also rather understandable, due to the visual similarity of the two. As explicated by Jill, such collection of impressive and funny examples helped her to better understand the capability and limitations of such CNN models, and think further about what they can do in the real-world contexts.

**Inferences of how the classification results were influenced by environmental factors**. Beyond getting a feel for the model's capabilities and limitations, examples shared by the students also revealed the inferences they made about how certain classification results were influenced by environmental factors (e.g., background, or adjacent objects). For example, Jill collected a case showing that how another object next to the target would influence the classification results: her hair clip was classified as a snorkel when put next to some cables. As she reasoned, the is because the cable looked similar to a snorkel tube and thus made the combination identified as snorkel. In another example, Ada found that her shoe was classified as more likely to be a sandal (56.5%) than a running shoe (31.2%). And as she inferred, this is due to that the color division of the shoe pattern (see Figure 3) made the model think that its upper part and lower part were two different objects. And the poor lighting of the environment also contributed to the confidence of classifying the shoe as a sandal. As shown above, the examples in this category suggest that the students' sense-making went beyond simply accumulating case-by-case knowledge of what worked and what did not with the CNN model. In many cases, they also tried to make inferences about why the CNN model generated certain (unexpected) results in given situations. Such inferences echoed what D.A. Schöne referred to as "theory-in-practice" [23], which manifested the practical understandings established by the students to interpret how the model worked in their environment.

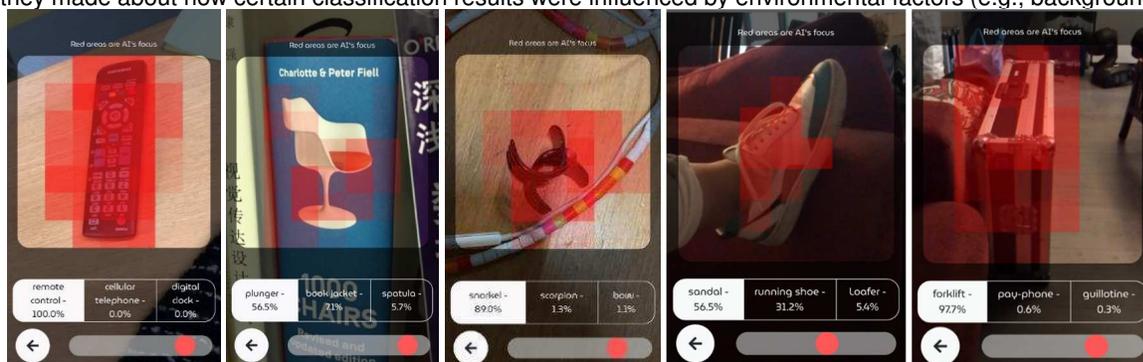

Figure 3. A few examples collected by the students when playing with their daily surrounding, to understand the functioning of the CNN model, from left to right: remote TV control (Jill), book (Sam), hair clip (Jill), sport shoe (Ada), suitcase (Ada).

**Experimentation on the spot with surrounding objects to test assumptions**. In addition to making inferences, there are also examples in which the students actively experimented with the CNN model to test their assumptions, e.g., by taking pictures of a target from different perspectives, or in combination with different objects. For instance, by experimenting with the mobile app, Ada found that the CNN model "doesn't seem to understand perspective". This was learnt by taking pictures of her suitcase from multiple perspectives: while the pictures that captured the front view (the largest surface) of the suitcase were correctly classified, the side view of her suitcase was mistakenly classified as a forklift. Another telling example is from Sam. He performed a series of experiments with his smartwatch, to test his assumptions about what features make the CNN model more confident that a target is a digital watch. His first hypothesis was that the appearance of a clock face screen would add to the confidence. To verify this, as Figure 4 shows, he took two pictures from the same perspective, which the only difference being showing/dimming the clock face screen. As a result, the one with the screen dimmed led to lower confidence (29.1%) than the other with the clock face screen (82.3%), Moreover, Sam hypothesized that being attached to an arm also contributes to the confidence of a target being a digital

Sam hypothesized that being attached to an arm also contributes to the confidence of a target being a digital watch. Hence, he did the next experiment by taking another picture of the smartwatch taken off his arm, from the similar perspective as the last pictures. The result supported his hypothesis: it was classified as a remote control instead of a digital watch. He further hypothesized that putting the watch on a cylinder object might also work like putting it on a real arm, so he attached the smartwatch to his bottle (see Figure 4), which indeed verified his assumption. As recognized by the students, such mentioned experimentations were another advantage of the mobile interface, since in the web interface, it would take more efforts to perform such controlled comparisons, Examples from this category therefore have shown that the mobile interface did not only supported the students' exploration and interpretation, but also enabled them to conduct on-the-spot experimentations to assess their practical assumptions about how the CNN model works.

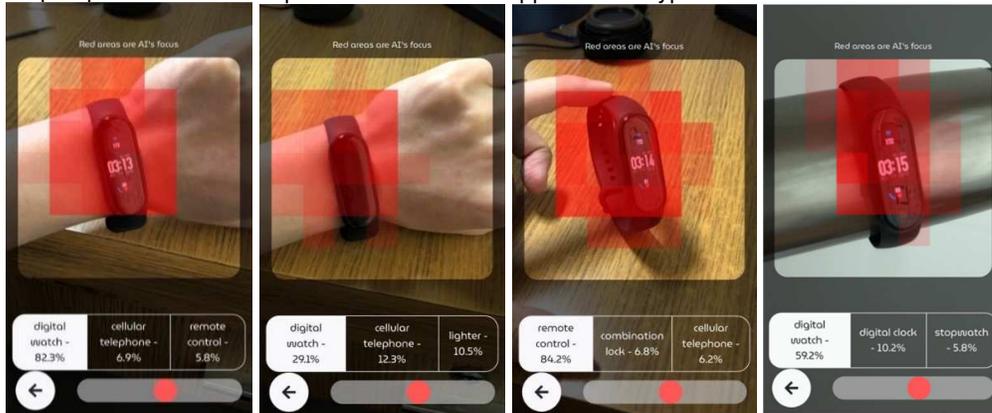

*Figure 4. On-the-spot experimentation conducted by Sam to test his hypotheses about what factors may increase the confidence of the pretrained CNN model recognizing his smartwatch, from left to right: with screen on and worn on the wrist; screen off worn on the wrist, screen on and taken off the wrist, screen on and put on a wrist-like object.*

## 6 CONCLUDING REMARKS

In this work, we present a mobile tool that enables nonexperts (design students as an example) to interactively make sense of a pretrained CNN model. Using a Class Activation Map approach, the tool intuitively visualizes for users how the CNN model makes certain predictions about a picture they take from their daily surroundings. The empirical findings confirmed its usefulness in terms of scaffolding nonexperts' practical understandings encompassing the functioning of the CNN model. Rich examples from the students' playful explorations surface three types of sensemaking processes that differ in the depths of thought. First, students used the tool to collect impressive or funny prediction results that helped them understand the capabilities and limitations of the model in real-world scenarios. Second, some (peculiar) results triggered the students' further inferences about how real-world environmental factors (e.g., lighting, perspective, adjacent object) could influence the prediction results. Lastly, the tool enabled the students to conduct on-the-spot experimentations to test their established understandings by creating variations in multiple pictures and comparing the results.

## REFERENCE

[1] M. Abadi *et al.*, "Tensorflow: A system for large-scale machine learning," in *12th {USENIX} symposium on operating systems design and implementation ({OSDI} 16)*, 2016, pp. 265–283.

[2] Y. Jia *et al.*, "Caffe: Convolutional Architecture for Fast Feature Embedding," in *Proceedings of the 22nd ACM International Conference on Multimedia*, 2014, pp. 675–678, doi: 10.1145/2647868.2654889.

[3] A. Paszke *et al.*, "Automatic differentiation in pytorch," 2017.

[4] K. Simonyan and A. Zisserman, "Very deep convolutional networks for large-scale image recognition," *3rd Int. Conf. Learn. Represent. ICLR 2015 - Conf. Track Proc.*, 2015.

[5] S. Mishra and J. M. Rzeszotarski, "Designing Interactive Transfer Learning Tools for ML Non-Experts," 2021, doi: 10.1145/3411764.3445096.

[6] Q. Yang, J. Suh, N.-C. Chen, and G. Ramos, "Grounding interactive machine learning tool design in how non-experts actually build models," in *Proceedings of the 2018 Designing Interactive Systems Conference*, 2018, pp. 573–584.

[7] M. Lin, Q. Chen, and S. Yan, "Network in network," *2nd Int. Conf. Learn. Represent. ICLR 2014 - Conf. Track Proc.*, 2014.

[8] M. D. Zeiler, G. W. Taylor, and R. Fergus, "Adaptive deconvolutional networks for mid and high level feature learning," *Proc. IEEE Int. Conf. Comput. Vis.*, pp. 2018–2025, 2011, doi: 10.1109/ICCV.2011.6126474.

[9] Q. Yang, A. Scuito, J. Zimmerman, J. Forlizzi, and A. Steinfeld, "Investigating how experienced UX designers effectively work with machine learning," in *Proceedings of the 2018 Designing Interactive Systems Conference*, 2018, pp. 585–596.

[10] Q. Yang, A. Steinfeld, C. Rosé, and J. Zimmerman, "Re-examining whether, why, and how human-ai interaction is uniquely difficult to design," in *Proceedings of the 2020 chi conference on human factors in computing systems*, 2020, pp. 1–13.

[11] G. Dove, K. Halskov, J. Forlizzi, and J. Zimmerman, "UX design innovation: Challenges for working with machine learning as a design material," in *Proceedings of the 2017 chi conference on human factors in computing systems*, 2017, pp. 278–288.

[12] R. Guidotti, A. Monreale, S. Ruggieri, F. Turini, D. Pedreschi, and F. Giannotti, "A survey of methods for explaining black box models," *arXiv*, vol. 51, no. 5, pp. 1–42, 2018.

[13] A. B. Arrieta *et al.*, "Explainable Artificial Intelligence (XAI): Concepts, taxonomies, opportunities and challenges toward responsible AI," *arXiv*, vol. 58, pp. 82–115, 2019.

[14] M. D. Zeiler, D. Krishnan, G. W. Taylor, and R. Fergus, "Deconvolutional networks," in *Proceedings of the IEEE Computer Society Conference on Computer Vision and Pattern Recognition*, 2010, pp. 2528–2535, doi: 10.1109/CVPR.2010.5539957.

[15] B. Zhou, A. Khosla, A. Lapedriza, A. Oliva, and A. Torralba, "Learning Deep Features for Discriminative Localization," *Proc. IEEE Comput. Soc. Conf. Comput. Vis. Pattern Recognit.*, vol. 2016-Decem, pp. 2921–2929, 2016, doi: 10.1109/CVPR.2016.319.

[16] R. R. Selvaraju, M. Cogswell, A. Das, R. Vedantam, D. Parikh, and D. Batra, "Grad-CAM: Visual Explanations from Deep Networks via Gradient-Based Localization," *Int. J. Comput. Vis.*, vol. 128, no. 2, pp. 336–359, 2020, doi: 10.1007/s11263-019-01228-7.

[17] D. Smilkov, N. Thorat, C. Nicholson, E. Reif, F. B. Viégas, and M. Wattenberg, "Embedding projector: Interactive visualization and interpretation of embeddings," *arXiv Prepr. arXiv1611.05469*, 2016.

[18] D. Linsley, D. Shiebler, S. Eberhardt, and T. Serre, "Learning what and where to attend," *7th Int. Conf. Learn. Represent. ICLR 2019*, pp. 1–21, 2019.

[19] Evgeny Demidov, "Interactive Heat map demo," 2019. https://www.ibiblio.org/e-notes/ml/heatmap.htm.

[20] D. Smilkov *et al.*, "Tensorflow. js: Machine learning for the web and beyond," *arXiv Prepr. arXiv1901.05350*, 2019.

[21] TensorFlow authors, "MobileNet," 2021. .

[22] M. Carney *et al.*, "Teachable machine: Approachable Web-based tool for exploring machine learning classification," in *Extended Abstracts of the 2020 CHI Conference on Human Factors in Computing Systems*, 2020, pp. 1–8.

[23] D. A. Schon, *The reflective practitioner: How professionals think in action*, vol. 5126. Basic books, 1984.